\documentclass[%
aps,
superscriptaddress,
bibnotes,
amsmath,amssymb,
prb,
floatfix,
longbibliography,
twocolumn,
]{revtex4-2}

\usepackage{float}
\usepackage{color}
\usepackage[usenames,dvipsnames,svgnames,table]{xcolor}
\usepackage[colorlinks=true,linkcolor=blue,urlcolor=blue,citecolor=blue]{hyperref}
\usepackage{graphicx}
\usepackage{dcolumn}
\usepackage{bm}
\usepackage{xcolor}
\usepackage{enumerate}
\usepackage{braket}
\usepackage[T1]{fontenc}
\usepackage{listings}
\usepackage{tabularx}
\usepackage{booktabs}
\usepackage{amsmath}
\usepackage[caption=false]{subfig}
\usepackage{ragged2e} 
\DeclareCaptionJustification{justified}{\justifying}

\usepackage[english]{babel}
\makeatletter
\def\bbl@set@language#1{%
  \edef\languagename{%
    \ifnum\escapechar=\expandafter`\string#1\@empty
    \else\string#1\@empty\fi}%
  \@ifundefined{babel@language@alias@\languagename}{}{%
    \edef\languagename{\@nameuse{babel@language@alias@\languagename}}%
  }%
  \select@language{\languagename}%
  \expandafter\ifx\csname date\languagename\endcsname\relax\else
    \if@filesw
      \protected@write\@auxout{}{\string\select@language{\languagename}}%
      \bbl@for\bbl@tempa\BabelContentsFiles{%
        \addtocontents{\bbl@tempa}{\xstring\select@language{\languagename}}}%
      \bbl@usehooks{write}{}%
    \fi
  \fi}
\newcommand{\DeclareLanguageAlias}[2]{%
  \global\@namedef{babel@language@alias@#1}{#2}%
}
\makeatother

\DeclareLanguageAlias{en}{english}

\newcommand{\br}{\bm{r}}
\newcommand{\bR}{\bm{R}}
\newcommand{\ubr}{\underline{\bm{r}}}
\newcommand{\ubR}{\underline{\bm{R}}}

\newcommand{\s}{_\mathrm{{\scriptscriptstyle S}}}
\newcommand{\h}{_\mathrm{{\scriptscriptstyle H}}}
\newcommand{\xc}{_\mathrm{{\scriptscriptstyle XC}}}

\newcommand{\electronic}{_\mathrm{e}}
\newcommand{\ionic}{_\mathrm{i}}

\newcommand{\bk}{\bm{k}}

\newcommand{\kB}{k_\mathrm{B}}

\begin{document}

\title{Machine learning the electronic structure of matter across temperatures}

\author{Lenz Fiedler}
\email{l.fiedler@hzdr.de}
\affiliation{Center for Advanced Systems Understanding (CASUS), D-02826 G\"orlitz, Germany}
\affiliation{Helmholtz-Zentrum Dresden-Rossendorf, D-01328 Dresden, Germany}

\author{Normand A.~Modine}
\affiliation{Sandia National Laboratories, Albuquerque, NM 87185, USA}

\author{Kyle D.~Miller}
\affiliation{Sandia National Laboratories, Albuquerque, NM 87185, USA}

\author{Attila Cangi}
\email{a.cangi@hzdr.de}
\affiliation{Center for Advanced Systems Understanding (CASUS), D-02826 G\"orlitz, Germany}
\affiliation{Helmholtz-Zentrum Dresden-Rossendorf, D-01328 Dresden, Germany}

\date{\today}

\begin{abstract}
We introduce machine learning (ML) models that predict the electronic structure of materials across a wide temperature range. Our models employ neural networks and are trained on density functional theory (DFT) data. Unlike most other ML models that use DFT data, our models directly predict the local density of states (LDOS) of the electronic structure. This provides several advantages, including access to multiple observables such as the electronic density and electronic total free energy.
Moreover, our models account for both the electronic and ionic temperatures independently, making them ideal for applications like laser-heating of matter.
We validate the efficacy of our LDOS-based models on a metallic test system. They accurately capture energetic effects induced by variations in ionic and electronic temperatures over a broad temperature range, even when trained on a subset of these temperatures. These findings open up exciting opportunities for investigating the electronic structure of materials under both ambient and extreme conditions.
\end{abstract}

\maketitle


\section{Introduction}
\label{sec:introduction}
Predicting the electronic structure of matter is essential for advancing scientific progress across various applications. Electronic structure calculations, which typically employ density functional theory (DFT)~\cite{HoKo1964, KoSh1965}, have become a routine tool in materials science and chemistry due to their accuracy and computational efficiency~\cite{Jo2015,PaJa2019}. 

However, as the demand for high-fidelity simulation data in emerging research areas increases, conventional DFT simulations face significant limitations. 
DFT calculations exhibit unfavorable scaling with both system size and temperature \cite{ofdftpaper}, limiting their applicability for current scientific challenges, particularly in studying materials under extreme conditions and within the warm dense matter regime \cite{POP2020, WDM1,WDM2}. Progress in this area not only contributes to the fundamental sciences by advancing our understanding of astrophysical objects \cite{PhysRevLett.81.5161,Militzer_2008, Daligault_2009,Nettelmann_2011}, but also propels technological developments by enabling the modeling of inertial confinement fusion capsule heating processes \cite{hu_2011_first}, radiation damage processes in reactor walls \cite{zarkadoula_2014_electronic, hammond2020_theoretical, zhou2021_enabling, sikorski_2023_machine}, and advanced manufacturing \cite{antony2018advanced,frazier2014metal}. Additionally, it supports diagnostics of scattering experiments conducted at free-electron laser facilities \cite{LCLS_2016, Tschentscher_2017} and promotes the emerging field of hot-electron chemistry for accelerating chemical reactions \cite{Mukherjee2013,Brongersma2015}.
A particularly relevant phenomenon in these applications involves rapidly driven electrons leading to transient non-equilibrium conditions resulting in hot electrons and cool nuclei which have also been observed in semiconducting and dielectric materials~\cite{harb_2008_electronically,li_2011_role}. 

To address these computational limitations, the electronic structure community has increasingly turned to machine learning (ML) techniques \cite{deepdive}. ML algorithms can accurately predict complicated relationships using tractable data samples. The application of ML to DFT has led to numerous approaches, with most focusing on predicting specific observables of interest \cite{wilkins_accurate_2019} or replacing DFT entirely with ML interatomic potentials (ML-IAPs), which capture the electronic total energy or total free energy landscape of a system and enable extended simulations of ionic dynamics \cite{wood_data-driven_2019, lukespaper}.
While existing ML-based approaches show promise in accurately predicting observables or capturing the energy landscape of quantum systems, most of them do not provide direct access to the electronic structure of matter. Knowledge of the electronic structure offers several advantages, e.g, the exploration of multiple observables beyond those targeted by a specific ML model. In recent work, the electronic structure has become a focus of ML models~\cite{BrVo2017,TsMi2020, MiRy2019}, albeit at a conceptual level. 

We have recently developed a practical formalism for modeling the electronic structure, represented by the local density of states (LDOS), using neural networks (NN) \cite{malapaper}. These models can reproduce multiple observables, such as the DOS, electronic densities, and electronic total free energies from the LDOS predictions. In our previous work~\cite{sizetransferpaper}, we demonstrated that these models can replace electronic structure calculations for systems larger than those accessible via conventional DFT. However, to provide ab-initio accuracy for advanced scientific applications, these models must also be able to make predictions across temperature ranges, since conventional DFT scales cubically with the temperature \cite{ofdftpaper} and retraining ML models for each temperature of interest is thus impractical. Applying such models across temperature ranges is promising since it is to be expected that accurate modeling of systems at high temperatures requires knowledge of the electronic structure, which our approach provides. 

The models discussed in this work are based on the formalism introduced in Ref.~\cite{malapaper} and account for electronic and ionic temperature separately. The ionic temperature enters the model through the ionic configurations sampled during molecular dynamics (MD) simulations and used as input to ML inference, while the electronic temperature is an explicit parameter in the expressions used to evaluate observables such as the electronic density and total free energy from the LDOS. As a result, our models can effectively handle situations where the electronic temperature surpasses the ionic temperature, such as in laser heating of matter. This characteristic of our ML model responds to a growing need for approaches that extend beyond conventional MD to tackle non-equilibrium conditions between electrons and nuclei, as recently developed within the framework of two-temperature MD ~\cite{darkins_2018_modelling, darkins_2018_simulating}.

Up until now, only a few studies have made efforts to incorporate electronic temperature into ML models trained to predict DFT data, with most studies only focusing on an explicit treatment of ionic temperature. Ref.~\cite{MLDFT_Temp1} develops an ML-IAP that directly addresses the electronic temperature, unlike typical ML-IAPs. Refs.~\cite{MLDFT_Temp2.2,MLDFT_Temp3} use an ML-approximated density of states (DOS), based on the ML models introduced in Ref.~\cite{MLDFT_Temp2}, to capture energetic effects associated with the electronic temperature and provide corrections to calculated observables. The framework outlined in Refs.~\cite{MLDFT_Temp2,MLDFT_Temp2.2,MLDFT_Temp3} shares similarities with the models discussed in this work. In both instances, the electronic temperature is treated independently from ionic temperature and enters the computation upon post-processing of ML predicted electronic structure quantities. Furthermore, both approaches exploit the property of the DOS to encode information about a system of interest at electronic temperatures higher than the DOS was originally computed, as detailed below. However, since the framework outlined by Refs.~\cite{MLDFT_Temp2,MLDFT_Temp2.2,MLDFT_Temp3} is confined to the DOS, it cannot access volumetric properties of interest, such as the electronic density or related energy contributions. Our models are distinct from both Ref.~\cite{MLDFT_Temp1} and Refs.~\cite{MLDFT_Temp2,MLDFT_Temp2.2,MLDFT_Temp3} since a spatially resolved representation of the electronic structure is predicted, which is then further processed to observables of interest at a specific electronic temperature. 

Within this work, we showcase the transferability and usefulness of our approach with a case study of aluminum as evidence that LDOS-based ML-DFT models can accurately predict the electronic structure of matter over a range of temperatures. To this end, we first investigate the influence of ionic and electronic temperatures on DFT simulations and LDOS based frameworks in Sec.~\ref{sec:resultsTheory}, before presenting and analyzing ML models trained on finite-temperature DFT simulations for aluminium at different temperatures in Sec.~\ref{sec:resultsModels}. The resulting analysis unveils that by directly predicting the LDOS, our models inherently incorporate thermal excitation effects. In contrast to conventional DFT simulations, our models mitigate the temperature scaling issue, requiring training data only at a few select temperatures to facilitate modeling of materials over a broad temperature range. This ability to interpolate across temperature regimes suggest that  LDOS-based ML-DFT models are particularly well-suited to investigating materials under extreme conditions. 

\section{Methods}
\subsection{Finite-Temperature DFT}
In the following, a brief outline of DFT simulations at finite-temperature, as they are employed for data generation in the context of temperature-transferable ML-DFT models, is provided. Generally, in electronic structure theory, a system of $L$ electrons and $N$ ions is typically described using collective electronic coordinates $\ubr=\{\br_1, ..., \br_L\}$ and ionic coordinates $\ubR=\{\bR_1, ..., \bR_N\}$, where $\br_j\in\mathbb{R}^3$ and $\bR_\alpha\in\mathbb{R}^3$. Within the framework of quantum statistical mechanics \cite{toda1983statistical}, thermodynamic properties of such a system can be obtained from the grand potential
\begin{equation}
    \Omega[\hat{\Gamma}] = \mathrm{Tr}(\hat{\Gamma}\hat{\Omega}) \; ,
    \label{eq:grandpotential}
\end{equation}
with the grand canonical operator
\begin{equation}
    \hat{\Omega} = \hat{H}-\mu \hat{N}-\kB {\tau\electronic} \hat{S} \,, \label{eq:grandcanonicaloperator}
\end{equation}
where $\hat{N}$ denotes the particle number operator, $\hat{S}$ the entropy operator, $\mu$ the chemical potential, and ${\tau\electronic}$ the (electronic) temperature. In the following, we will distinguish between the ionic temperature ${\tau\ionic}$ and the electronic temperature ${\tau\electronic}$. In Eq.~(\ref{eq:grandcanonicaloperator}), $\hat{H}$ denotes the Hamiltonian that represents all interactions between electrons and ions. In electronic structure theory, the Born-Oppenheimer approximation \cite{born_zur_1927} is typically employed, resulting in the Born-Oppenheimer Hamiltonian
\begin{equation}
    \hat{H} =\hat{T}+\hat{V}^{ee}+\hat{V}^{ei} + \hat{V}^{ii} \; , \label{eq:Hamiltonian_full}
\end{equation}
where $\hat{T}=\sum_{j}^{L} -\nabla_j^2/2$ denotes the kinetic energy operator of the electrons, $\hat{V}^{ee}=\sum_{j}^{L} \sum_{l \ne j}^{L} 1/(2\vert\br_j-\br_l\vert)$ the electron-electron interaction, $\hat{V}^{ei}= -\sum_{j}^{L} \sum_{\alpha}^{N} Z_\alpha/\vert\br_j-\bR_\alpha\vert$ the electron-ion interaction (with the $\alpha$-th ion having the charge $Z_\alpha$), and $\hat{V}^{ii}=\sum_{\alpha}^{N} \sum_{\beta \ne \alpha}^{N} Z_\alpha Z_\beta/(2\vert\bR_\alpha-\bR_\beta\vert)$ the ion-ion interaction. The Born-Oppenheimer approximation assumes that since the mass of the ions far exceeds that of the electrons, the electrons can be assumed to reach thermal equilibrium with each other on a comparatively small time scale. Thus, the kinetic energy of the ions and the ion-ion interaction can be treated classically and contribute simple additive terms to the energy. The resulting equations thus only depend on the ionic positions parametrically, and we will omit the $\ubR$ dependence in the following.

The statistical density operator $\hat{\Gamma}$ in Eq.~(\ref{eq:grandpotential}) is used to compute statistical averages. It is defined as a sum over all $L$-particle eigenstates $\Psi_{L,j}$ of the Born-Oppenheimer Hamiltonian as   
\begin{equation}
    \hat{\Gamma} = \sum_{L,j} w_{L,j} \ket{\Psi_{L,j}}\bra{\Psi_{L,j}} \; , \label{eq:statisticaldensity}
\end{equation}
where $w_{L,j}$ denote statistical weights that satisfy the normalization condition $\sum_{L,j} w_{L,j}=1$. In the grand canonical ensemble, thermal equilibrium is defined as the statistical density operator that minimizes the grand potential.  

The framework outlined here does not provide a practical approach for performing large-scale electronic structure calculations due to the electron-electron interaction in Eq.~(\ref{eq:Hamiltonian_full}) and the dimensionality of $\Psi_{L,j}$. Instead, DFT based on the Hohenberg-Kohn theorems \cite{HoKo1964}, their generalization to finite temperatures \cite{mermin_thermal_1965}, and the Kohn-Sham scheme \cite{KoSh1965} is used to describe the grand potential as a functional of the electronic density $n$. Specifically, the grand potential can be expressed as
\begin{align}
    \Omega[n] = E[n] -\kB {\tau\electronic} S\s[n] -\mu L \; , \label{eq:grandpotential_density}
\end{align}
with the electronic total energy
\begin{equation}
    E[n] = T\s[n]+E_\mathrm{H}[n]+E_\mathrm{XC}[n]+E_{ei}[n]+E_{ii} \label{eq:totalenergy} \; ,
\end{equation}
where the electronic density is restricted to densities corresponding to those stemming from many-body wavefunctions $\Psi_{L,j}$. Here, $T\s[n]$ denotes the kinetic energy of the auxiliary system of non-interacting fermions, $S\s[n]$ the non-interacting electronic entropy, $E_\mathrm{H}[n]$ the Hartree energy, which is the classical electrostatic interaction energy, $E_{ei}[n]$ the electron-ion interaction, which reduces to the interaction of the electronic density with an external potential, while $E_{ii}$ refers to the constant shift in energy due to ion-ion interaction. $E_\mathrm{XC}[n]$ refers to the exchange-correlation energy, which incorporates any electronic contributions not captured by the aforementioned energy terms, and which needs to be approximated in practice. Consequently, the accuracy of DFT calculations chiefly depends on appropriate approximations for this term. 
A plethora of practical approximations exist, such as the local density approximations (LDA, e.g.~PW91 \cite{KoSh1965, ceperley_ground_1980}) or generalized gradient approximations (GGA, e.g.~PBE \cite{perdew_accurate_1986,perdew_accurate_1992,perdew_generalized_1996}).
Furthermore, the XC functional should depend explicitly on temperature \cite{KV83}. However, this explicit temperature-dependence is often unclear and usually omitted in standard calculations. Instead, the temperature dependence is crudely included through the density in thermal equilibrium. Recent advances \cite{PPFS11,DT11,PPGB14,DT16,JB16,SJB16,BSGJ16,SSB18,SB20,MDBV21,MDVC22,MLVC23,MLVCb23} have reignited the development of temperature-dependent XC approximations. Studies of the electron liquid \cite{PD00} and uniform electron gas \cite{GR80,DT81,LM83,SD13} have aided in the construction of local \cite{BCDC13,KSDT14,DGSMFB16,GDSMFB17} and generalized gradient approximations \cite{SD14,KDT18} to the temperature-dependent XC contribution.

The subscript $\mathrm{S}$ in Eq.~(\ref{eq:grandpotential_density}) indicates that the kinetic energy and the electronic entropy are usually calculated via the Kohn-Sham ansatz \cite{KoSh1965}. This ansatz employs an auxiliary system of non-interacting single-particle wavefunctions $\phi_j$, which are constrained to reproduce the interacting electronic density through 
\begin{equation}
    n(\br, \ubR)=\sum_j f^{\tau\electronic}(\epsilon_j)|\phi_j|^2 \; , \label{eq:density}
\end{equation}
where $f^{\tau\electronic}$ denotes the Fermi-Dirac distribution, while $\phi_j$ and $\epsilon_j$ are the eigenfunctions and eigenvalues of the Kohn-Sham equations
 \begin{equation}
     \left[-\frac{1}{2} \nabla^2 +v\s(\br)\right]\phi_j(\br) = \epsilon_j\phi_j(\br) \; , \label{eq:KSequation}
 \end{equation}
where $\phi_j(\br)$ are referred to as Kohn-Sham wave functions or orbitals. The number of Kohn-Sham eigenstates and eigenvalues to be considered is $L$ for calculations at $\tau\electronic=0\mathrm{K}$, i.e., the sum in Eq.~(\ref{eq:density}) runs from $j=1$ to $j=L$. However, in the finite temperature picture, one has to account for thermal excitations of the electrons by calculating $\eta$ additional eigenstates, where $\eta$ is chosen such that $f^{\tau\electronic}(\epsilon_j)$ is negligible for $j > L + \eta$. As one moves to higher temperatures, $\eta$ has to be increased, which causes the unfavorable scaling of DFT with temperature. 

The Kohn-Sham potential $v\s$ is determined self-consistently to reproduce the interacting density via Eq.~(\ref{eq:density}). Note that several terms within the Kohn-Sham framework implicitly depend on  ${\tau\electronic}$, such as the eigenstates and eigenfunctions of Eq.~(\ref{eq:KSequation}), which are not explicitly denoted here for readibility.

With this finite-temperature DFT (FT-DFT) framework, which is further discussed in-depth in Ref.~\cite{PPGB14}, practical calculations at arbitrary temperatures become feasible. All DFT simulations performed in the context of this work were performed at finite temperatures using the framework outlined above. To further connect FT-DFT with dynamical studies, one obtains the electronic total free energy with the FT-DFT framework via Eq.~(\ref{eq:grandpotential_density}) as
\begin{equation}
    A[n] = \Omega[n]+\mu L= E[n] - \kB\tau\electronic S\s[n]\; , \label{eq:totalfreeenergy} 
\end{equation}
which enables calculating the force acting on the $\alpha$-th ion $\bm{F}_\alpha=\partial A / \partial \bR_\alpha$. These forces can be used, e.g., in MD simulations, which yield thermodynamic observables. Here, the ionic temperature ${\tau\ionic}$ is once again relevant and controlled via thermostats, such as the the Nosé-Hoover thermostat \cite{hoover,nose}. These thermostats ensure that thermodynamic sampling is performed in the correct thermodynamic ensemble, such as a canonical ensemble of ions, where $N$, ${\tau\ionic}$ and the volume of the simulation cell $V$ is held constant.

\subsection{DFT machine learning model}
\label{sec:mlmodels}
While finite-temperature DFT enables practical calculations of many systems, it still has inherent computational scaling limitations. The standard DFT approach scales with $N^3$, making it very difficult to simulate systems involving more than a few thousand atoms. While there are techniques that reduce this scaling behavior to $N$, such as orbital-free DFT \cite{LiCa2005} or linear-scaling DFT \cite{Ya1991,GoCo1994} through appropriate approximations, neither route enables a general replacement of KS-DFT simulations. 

As discussed above, more Kohn-Sham wavefunctions need to be included in Eq.~(\ref{eq:KSequation}) at higher temperatures, leading to an additional computational overhead that scales unfavorably with temperature \cite{ofdftpaper}. These limitations pose a challenge for generating first-principles data for practical applications, particularly for increasingly relevant investigations into matter under extreme conditions. As a result, ML is emerging as a promising route for overcoming these limitations.

ML is based on algorithms that can improve their performance through observed data, i.e., they can learn \cite{mitchell_machine_1997}. By training models on a representative set of data, it is possible to make predictions in a fraction of the time it would take to perform original data collection with conventional algorithms. In the context of DFT, ML models can be trained on a number of potentially costly simulations and thereafter replace the need for DFT simulations, leading to drastically reduced simulation times. 

There are various ML approaches for DFT simulations that differ widely in their goals and application purposes. In Ref.~\cite{deepdive}, we have identified common avenues of research, including ML models that directly predict quantities of interest for a subset of components (e.g., Ref.~\cite{pedersen_high-entropy_2020,ye_deep_2018,wood_data-driven_2019}) and models that predict total (free) energies and atomic forces, known as machine-learned interatomic potentials (ML-IAPs, e.g., Ref.~\cite{gartner_signatures_2020, deringer_machine_2017,wilkins_accurate_2019}). The latter can easily be integrated into MD frameworks, replacing DFT simulations as the primary engine for extended dynamical simulations. IAPs built on (semi-)empirical approximations have been used in this capacity before the advent of ML. However, ML-IAPs generally provide an even better reproduction of the electronic total free energy surface of a system of interest, allowing for highly accurate thermodynamic sampling of observables \cite{eval_mlips}. 

Current ML approaches for DFT calculations are limited in the quantities they predict and do not provide the full electronic structure of simulated systems. We have recently developed an ML-DFT framework \cite{malapaper,hyperparameterpaper} that overcomes this issue. It provides a spatially-resolved representation of the electronic structure within the local neighborhood of each point in real-space, $\br$, for systems of arbitrary size by predicting the LDOS. Based on this representation of the electronic structure, other, fundamental observables of interest (such as the electronic density) may be accessed. The feasibility of such a local description of the electronic structure is based on the assumption of nearsightedness of the electronic structure \cite{nearsightedness}. The LDOS is defined as
\begin{equation}
    d(\epsilon, \br) = \sum_j |\phi_j(\br)|^2 \delta(\epsilon-\epsilon_j) \label{eq:ldos}
\end{equation}
and, from a computational point of view, is a vector in the energy domain $\epsilon$ for each point in real-space $\br$. Corresponding to the Kohn-Sham system as discussed above, the upper boundary of the summation in Eq.~(\ref{eq:ldos}) depends on the number of Kohn-Sham eigenstates sampled, i.e., $L+\eta$. In order for models to be transferable in the $\tau\electronic$ domain, we choose $\eta$ such that unoccupied Kohn-Sham orbitals are included in the LDOS as well; thus, the same LDOS can be used at higher electronic temperatures to accurately compute energies. 

Through the LDOS, the electronic density and the density of states $D$ are obtained as
\begin{align}
    n(\br) &= \int d \epsilon f^{\tau\electronic}(\epsilon)d(\epsilon, \br) \; , \label{eq:density_from_ldos}\\
    D(\epsilon) &= \sum_j \delta(\epsilon-\epsilon_j) = \int d \br d (\epsilon, \br) \,, \label{eq:dos}
\end{align}
which are both of direct interest for electronic structure theory. Moreover, the electronic total energy in Eq.~(\ref{eq:totalenergy}) and the electronic total free energy in Eq.~(\ref{eq:totalfreeenergy}) can be expressed solely in terms of $d$ as
\begin{align}
    E[d] = &E_b[D[d]] + E_\mathrm{XC}[n[d]] - E\h[n[d]] \label{eq:totalenergy_ldos} \; , \\  \nonumber &-\int d \br v\xc(\br)n[d](\br) \; ,\\
    A[d] = &E[d] -\kB{\tau\electronic} S\s[D[d]] \; , \label{eq:totalfreeenergy_ldos}
\end{align}
where $E_b = \int d\epsilon f^{\tau\electronic} (\epsilon)\epsilon D(\epsilon)$ denotes the band energy and $v\xc$ the exchange-correlation potential. For a complete derivation of this framework refer to Ref.~\cite{malapaper}. Consequently, if the LDOS of a system can be predicted accurately, it becomes possible to determine both the electronic structure and the energetics of the system, and automatic differentation can be used to obtain the atomic forces. 

The LDOS can be predicted by a model $M[\lambda]$ in the form of 
\begin{equation}
    \Tilde{d}(\epsilon, \br) = M(B(J, \br)[\ubR])[\lambda] \; , \label{eq:prinicipalmapping}
\end{equation}
where $\lambda$ represents hyper-parameters that describe both the model (i.e., type of ML algorithm, characteristic features) and the fitting techniques (i.e, training parameters of the model, data employed), and for which we have developed techniques for rapid optimization \cite{hyperparameterpaper}. In our ML models, we use neural networks (NN) for the actual ML task. NNs can learn complicated relationships between sets of data through nested linear transformations and non-linear activation functions, based on individual units called artificial neurons or perceptrons \cite{minsky_perceptrons_1987}. The output of the $l+1$-th layer of an NN is calculated from the outputs of the $l$-th layer as
\begin{equation}
    \bm{y}^{l+1} = \sigma(\mathbf{W}^l\bm{y}^l+\bm{b}^l) \; ,
\end{equation}
where $\mathbf{W}^l$ and $\bm{b}^l$ are the weights and biases of the $l$-th layer, which are the tuneable parameters, and $\sigma$ is a non-linear activation function. The process of determining the optimal $\mathbf{W}$ and $\bm{b}$ is known as training~\cite{nielsen_neural_2015}. 

The first layer of the NN receives input data, which in the case of our ML models are descriptors with dimensionality $J_\mathrm{max}$ with $J=1,...,J_\mathrm{max}$, denoted as $B$ in Eq.~(\ref{eq:prinicipalmapping}). These descriptors capture information about the ionic structure locally around each point in space $\br$. The locality of this mapping is essential for scalability; since $M$ learns to predict the electronic structure at each point $\br_p$ independently of distant points $\br_q$, the model can be applied to diverse or large-scale systems as long as the observed descriptors $B$ are close to those in the training set. 

We employ a grid-point generalization of bispectrum descriptors \cite{thompson_spectral_2015} to encode the local ionic structure, as described in Refs.~\cite{malapaper, hyperparameterpaper}. The bispectrum descriptors at point $\br$ represent the atomic density around $\br$ in terms of a basis set expansion. 

\begin{figure}[htp]
    \centering
    \includegraphics[width=0.95\columnwidth]{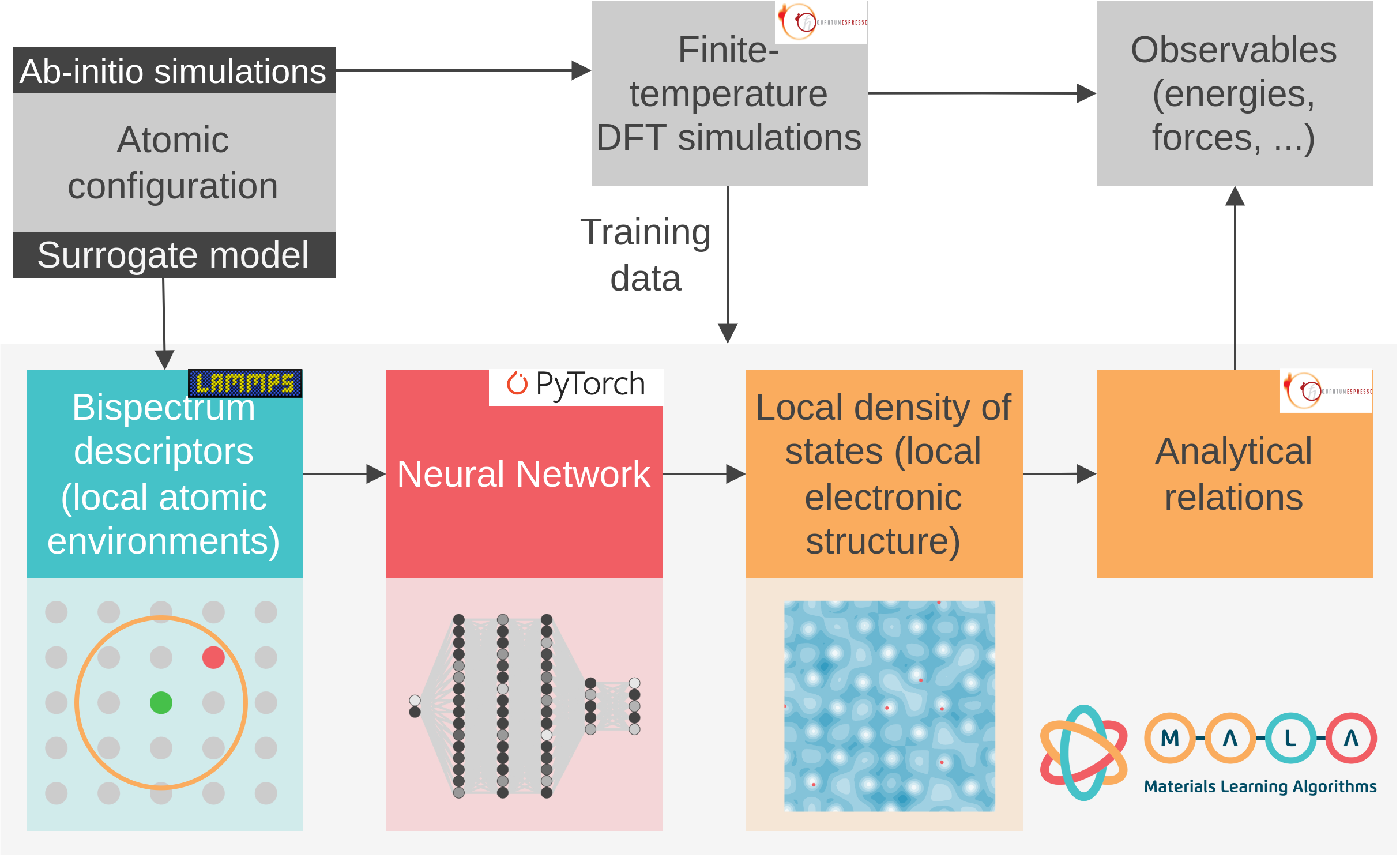}
    \caption{Overview of the MALA framework and the open source libraries used for constructing the full ML pipeline. The pictograms below depict the individual steps of the framework, which involves calculating local descriptors at an arbitrary grid point (green) from atoms (red) within a certain cutoff radius (orange), a neural network, and the electronic structure, specifically the electronic density for a cell of aluminum atoms (red). The pictograms are adapted from Ref.~\cite{sizetransferpaper}.}
    \label{fig:malaworkflow}
\end{figure}

This framework is transferable to different system sizes \cite{sizetransferpaper} and is applicable to a wide range of systems, as long as the LDOS can be accurately calculated via DFT for model training. We have developed an open-source software package called \textit{\textbf{Ma}terials \textbf{L}earning \textbf{A}lgorithms} (MALA) \cite{malagithub} which implements this LDOS based framework and interfaces with popular open-source libraries such as Quantum ESPRESSO, LAMMPS, and PyTorch. Fig.~\ref{fig:malaworkflow}, adapted from Ref.~\cite{sizetransferpaper}, illustrates the framework. Additionally, the transferability of models extends to different temperature ranges, which will be discussed in the following. 

\subsection{Computational details}
\label{sec:compdetails}
We investigate the temperature transferability of ML-DFT models using an aluminum data set covering ionic temperatures from 100K to 933K. The 933K data corresponds to the solid phase and has been previously used in Ref.~\cite{malapaper}. For other temperatures, ionic configurations were sampled from DFT-MD trajectories generated using the VASP code \cite{kresse_ab_1993,kresse_efficient_1996, kresse_efficiency_1996}, employing a $2\times2\times2$ Monkhorst-Pack \cite{monkhorst_special_1976} $\bk$-grid, a plane-wave basis set with a cutoff energy of 500 eV, a PAW pseudopotential \cite{blochl_projector_1994,kresse_ultrasoft_1999} and the PBE \cite{perdew_accurate_1986,perdew_accurate_1992,perdew_generalized_1996} exchange-correlation functional. Further, the DFT-MD trajectories were calculated with a time step of 2 $fs$, and in order to select uncorrelated configurations, every 147th step was sampled for subsequent DFT and LDOS calculations. This number was chosen empirically but based on the intuition that most high frequency phonons, which are strongly related to the local ionic structure and exhibit frequencies of multiple $THz$ in the case of aluminium \cite{aluminiumphononrelation}, should be able to perform a few full vibrations between sampling individual configurations.

The LDOS was then calculated for these ionic configurations using the Quantum ESPRESSO code \cite{giannozzi_q_2020,giannozzi_quantum_2009,giannozzi_advanced_2017} with computational parameters consistent with those in Ref.~\cite{malapaper}. Quantum ESPRESSO has been employed for data generation since calculation of the LDOS requires an adaptable width for the approximation of the $\delta$ distribution with Gaussians in Eq.~(\ref{eq:ldos}), as detailed in Ref.~\cite{malapaper}. The open source nature of the Quantum ESPRESSO code allowed for the necessary changes to be implemented into the code, and such capability has been made publicly available in Quantum ESPRESSO version 7.2. For a subset of the configurations, we also performed calculations at electronic temperatures that differ from ionic temperatures. This was done to investigate effects of the electronic temperature, as shown in Sec.~\ref{sec:resultsTheory}. For actual model training, which is described in Sec.~\ref{sec:resultsModels}, the same electronic and ionic temperatures were consistently used. The full dataset can be found in Ref.~\cite{ellis_ldossnap_2021}. 

MALA version 1.2.0 \cite{malagithub} was used to build and train the ML models, and the model parameters were kept consistent with the 933K model described in Ref.~\cite{ellis_ldossnap_2021}, with the exception being the second set of models calculated for Fig.~\ref{fig:temptransfer_superhybrid}, where the layer width was increased from 4000 neurons to 6000 neurons. Models and training scripts can be obtained via Ref.~\cite{fiedler_lenz_2023_2266}. Accessing individual data points in a randomized order is essential when training neural networks, and it is a standard practice in this field. However, for the volumetric data used in our training scheme, loading all the relevant data files into memory at the same time was not feasible due to their size. Instead "lazy loading" was employed, i.e., volumetric data corresponding to one ionic configuration at a time was loaded into memory. Doing so prevents the training code from properly randomizing data across ionic configurations during the training process, and one only randomizes the order at which ionic configurations are accessed. For earlier applications of this framework, such as Ref.~\cite{malapaper}, this did not constitute a problem, since the individual ionic configurations exhibit a comparatively low variance in terms of local ionic and electronic environments. However, for the higher variance observed for configurations sampled across multiple temperatures, this lack of fully randomized access initially led to accuracy issues for ML models. To combat these issues, we used separate scripts to read data and mix relevant volumetric data files into new sets of randomized files to ensure randomized access to data points for each training experiment. The relevant scripts for this can be found in Ref.~\cite{fiedler_lenz_2023_2266}.

\section{Learning the electronic temperature}
\label{sec:resultsTheory}
The central result of this work are LDOS based ML-DFT models that explicitly take the electronic temperature into account. This constitutes a fundamental difference to most other established ML-DFT frameworks, which only consider ionic temperature. Naturally, this brings up the question of the orders of magnitude for energy effects arising from both ${\tau\ionic}$ and ${\tau\electronic}$, since changes in either temperature are expected to have noticeable impact on total free energies. It is important to investigate the magnitudes and relative behaviors of these changes in energy, as the ML models discussed here aim to recover these effects. Thus, in the following, the influence of electronic and ionic temperature are investigated from a theoretical point of view. 

Established frameworks for learning the energy of materials, known as ML-IAPs take a set of ionic positions, $\ubR$, and predict the energy and often forces which are the derivative of the energy with respect to the positions. Usually, these models learn the relative energy with respect to some reference configuration (e.g., isolated atoms) rather than total energies or total free energies.  The ionic temperature ${\tau\ionic}$, which can be regulated though thermostats in dynamic simulations, determines the distributions of ionic velocities and positions.  Even though ML-IAPs are not directly dependent on the ionic temperature, they can learn to correctly predict dynamics for a range of temperatures once trained on ionic configurations representative of those temperatures.

Revisiting Eq.~(\ref{eq:grandpotential_density}) reveals that the Kohn-Sham electronic total free energy depends on the electronic temperature ${\tau\electronic}$, both directly through the electronic kinetic energy and entropy terms as well as implicitly through the electronic density, which is calculated via Eq.~(\ref{eq:density}) with the Fermi-Dirac distribution. Thus, for a given set of ionic positions, the Kohn-Sham electronic total free energy can have a whole range of values depending on the electronic temperature. In contrast, conventional ML-IAPs assume that the energy depends only on ionic positions, and thus it is impossible for them to capture the effects of different electronic temperatures.

The LDOS models discussed in Sec.~\ref{sec:mlmodels} address this shortcoming of regular ML models for finite-temperature DFT by learning the LDOS instead of directly learning the energy. Specifically, the electronic total free energy is obtained from the LDOS using Eq.~(\ref{eq:totalfreeenergy_ldos}), where the direct dependence of the electronic kinetic energy, entropy, and density on $\tau\electronic$ is taken into account.  Note that the LDOS $d(\epsilon, \br)$ also depends on ${\tau\electronic}$ implicitly through the Kohn-Sham eigenvalues and eigenfunctions, which in turn depend on the self-consistently determined Kohn-Sham potential.  In principle, $\tau\electronic$ could be added to the descriptors used to calculate $d(\epsilon, \br)$ and a suitable set of ${\tau\electronic
}$-dependent training data could be used to capture the dependence of the LDOS on ${\tau\electronic}$.  Instead, we will investigate the simple approximation in which the training data for the LDOS ML model is generated with $\tau\electronic = \tau\ionic$, and the ML model does not include any explicit dependence on ${\tau\electronic}$.  We will find that this approximation captures the great majority of the energetic effects of changing ${\tau\electronic}$.

\begin{figure}[h]
    \centering
    \includegraphics[width=0.95\columnwidth]{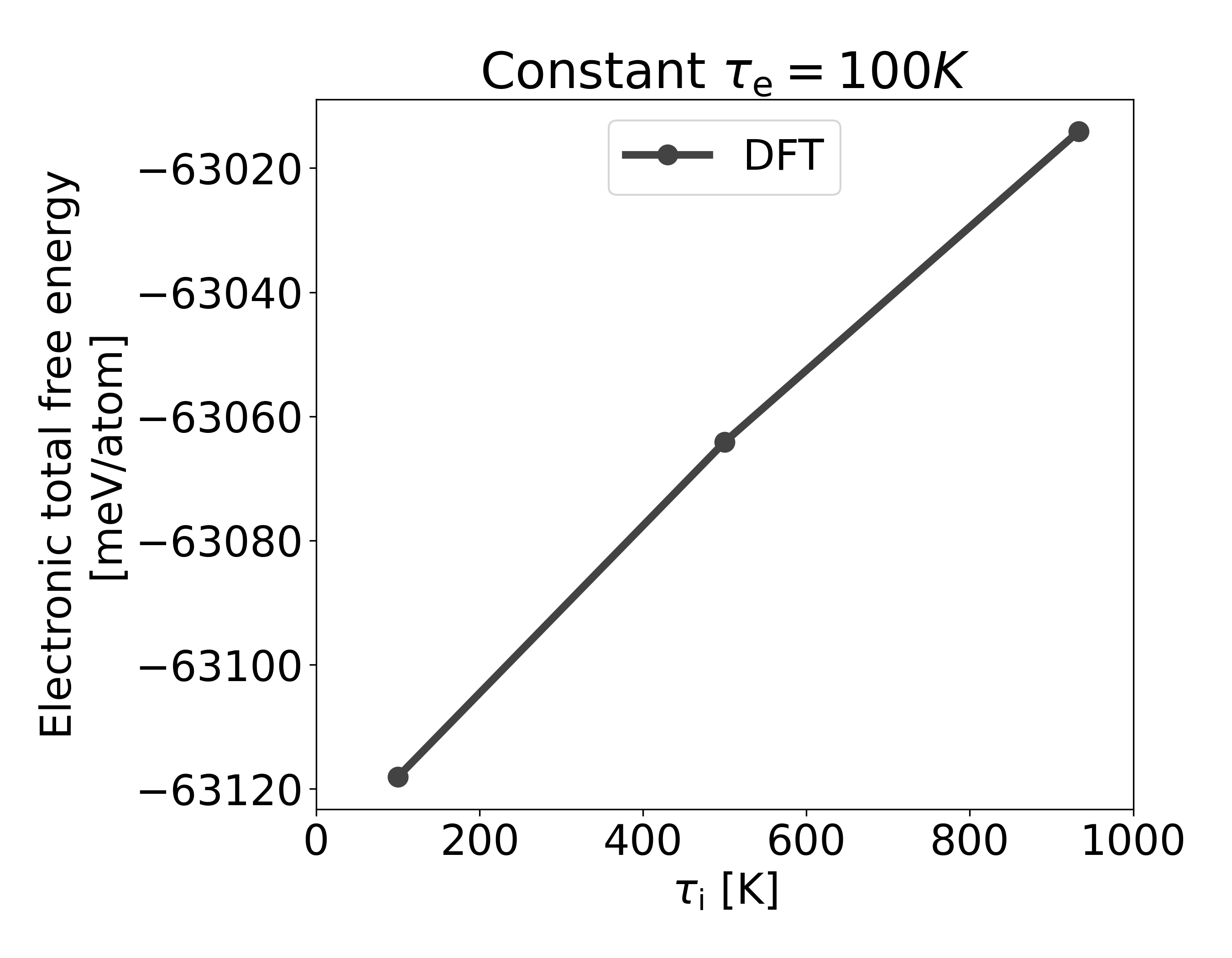}
    \caption{Changes in DFT energy when keeping electronic temperature constant while varying ionic temperature. For each point depicted, the electronic total free energy of ionic configurations sampled at the given ionic temperature has been calculated and averaged.}
    \label{fig:ionic_temperature_comparison}
\end{figure}

\begin{figure*}[h]
    \centering
    \includegraphics[width=0.95\textwidth]{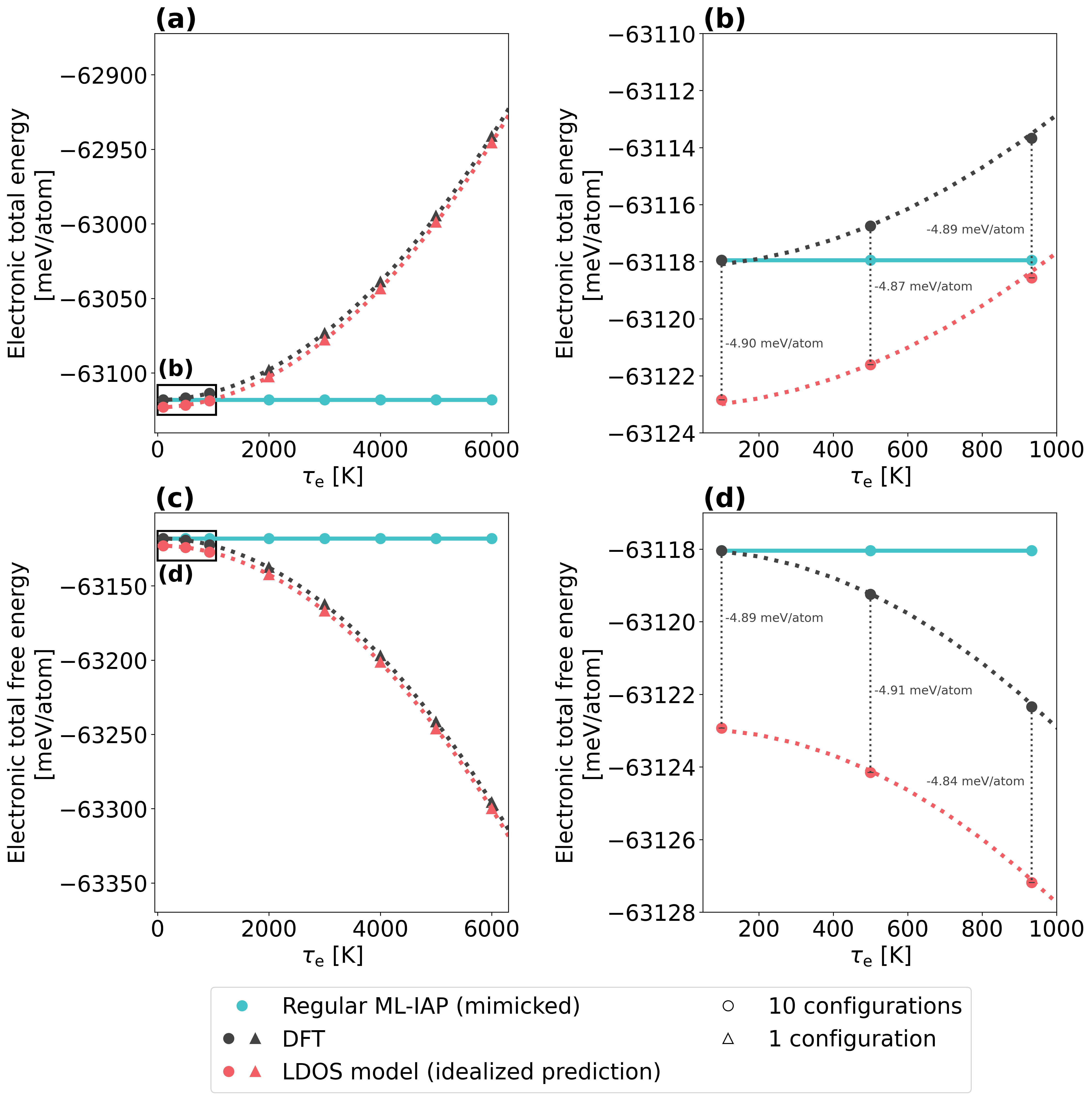}
    \caption{Changes in electronic total and total free energy as defined by Eqs.~(\ref{eq:totalenergy}), (\ref{eq:totalfreeenergy}), (\ref{eq:totalenergy_ldos}) and (\ref{eq:totalfreeenergy_ldos}) when keeping ionic temperature constant while varying electronic temperature. Ionic temperature was kept constant by sampling ionic configurations at 100K and not changing the ionic structure thereafter. Data was collected either by averaging energies over 10 ionic configurations (circles) or selecting one ionic configuration (for the temperature range above 933K, shown as triangles). The dotted lines are quadratic fits to the entire data set. In \textbf{(b)} and \textbf{(d)}, a magnification of the results given in \textbf{(a)} and \textbf{(c)} are shown, to give more detail for the training range of the ML models and energetic differences between LDOS and wave function based energy equations. For the grey curves, regular DFT calculations with varied electronic temperatures were performed. The blue curves show the expected behavior of a regular ML-IAP, which reproduces the same energies for all electronic temperatures, since the same ionic configurations are used. The red curve show the best case scenario for an LDOS based model - here, the LDOS sampled at 100K has been evaluated at higher electronic temperatures using Eq.~(\ref{eq:totalenergy_ldos}) and (\ref{eq:totalfreeenergy_ldos}).}
    \label{fig:electronic_temperature_comparison}
\end{figure*}

To quantify the influences of ionic and electronic temperature, we have performed DFT calculations in which ${\tau\ionic}$ or ${\tau\electronic}$ was kept constant while the other temperature was varied. These calculations were conducted for 256 aluminum atoms at 100K, 500K, and 933K (melting point, solid phase) for the ionic temperature experiments; for the electronic temperature experiments, temperatures up to 6000K were investigated. The results of this investigation are shown in Fig.~\ref{fig:ionic_temperature_comparison} and Fig.~\ref{fig:electronic_temperature_comparison}.

Fig.~\ref{fig:ionic_temperature_comparison} shows the impact of the ionic temperature. Here the electronic total free energy has been calculated for ten ionic configurations sampled at $\tau\ionic=[100\mathrm{K},\,500\mathrm{K},\,933\mathrm{K}]$ with DFT and a constant $\tau\electronic=100\mathrm{K}$. Afterward, the energies have been averaged per temperature. As expected, sampling ionic configurations at different ionic temperatures results in a change in energy, even when the electronic temperature is kept constant. Fig.~\ref{fig:ionic_temperature_comparison} gives an insight into how large this change in energy is to be expected for changes in the ionic temperature. Here, we observe an energy change of slightly more than 100 meV/atom for the range of 100K to 933K. This result agrees well with the classical Harmonic heat capacity, which would give a $\frac{3}{2} k_{B} (T_1 - T_0) = 108$ meV/atom increase in potential energy for a temperature change from $T_0 = 100K$ to $T_1 = 900K$.

In contrast, Fig.~\ref{fig:electronic_temperature_comparison} shows the behavior of the electronic total free energy and electronic total energy when electronic temperature is varied (both in black), while ionic temperature is kept constant. As mentioned above, one advantage of the models discussed here is their ability to treat electronic and ionic temperature independently, which is an inherent capability of DFT simulations not easily reproduced in ML-DFT frameworks. This property becomes relevant, e.g., when electrons are heated to very high temperatures by a laser source and only subsequently heat the cold ions. In the case of aluminium, one can estimate the electronic temperature at which hot electrons are able to melt solid aluminium at low temperatures at roughly 5800K (see App.~\ref{app:al_motivation}). We thus investigate electronic temperatures up to 6000K in Fig.~\ref{fig:electronic_temperature_comparison} and further show ML inference results for this temperature range in Sec.~\ref{sec:resultsModels}. 

The black curve in Fig.~\ref{fig:electronic_temperature_comparison} was calculated using ionic configurations sampled at ionic temperature 100K and DFT simulations with increasing electronic temperature. It can clearly be seen that overall, one observes a quadratic dependence for both energies w.r.t.~electronic temperature. While electronic total energy increases, the electronic total free energy decreases, due to an increase of the electronic entropy term, which is included in the electronic total free energy according to Eq.~(\ref{eq:totalfreeenergy}) and is negative in sign. It should be noted that for the temperature range of 100K to 933K, on which the models discussed in Sec.~\ref{sec:resultsModels} were originally trained, the associated energy change is of the magnitude of a few meV/atom, and thus smaller than for changes in ionic temperature within the same range.

Additionally, Fig.~\ref{fig:electronic_temperature_comparison} shows the hypothetical behavior of ML models for changes in electronic temperatures. There, the blue curve corresponds to the (mimicked) behavior of a regular ML-IAP. Since ML-IAPs generally only operate on ionic configuration data, any such model has no concept of electronic temperature, and thus its energy predictions would be entirely unaffected by changes in electronic temperature, yielding a constant prediction. Naturally, for an actually trained ML-IAP, this line may be shifted upward or downward, based on the training routines; the relative behavior would be the same, however. 

Conversely, the red curve in Fig.~\ref{fig:electronic_temperature_comparison} shows the effects of using a $\tau\electronic$ independent LDOS to predict electronic total energies and total free energies over a range of electronic temperatures.  Here, we take the LDOS calculated at $\tau\electronic=100\mathrm{K}$ for each ionic configuration and predict electronic total (free) energies by substituting $\tau\electronic$ in  Eq.~(\ref{eq:density_from_ldos}), (\ref{eq:totalenergy_ldos}) and (\ref{eq:totalfreeenergy_ldos}). Ideally, this represents the results that would be obtained from a LDOS based ML model without any explicit dependence on $\tau\electronic$ in the LDOS prediction.  It can be seen that one almost perfectly recovers the relative behavior of the electronic total (free) energies with this approximation.

One might be puzzled by the approximately -4.9 meV energy offset between the DFT energies and the LDOS-based energies.  As has been discussed in Ref.~\cite{malapaper} and \cite{sizetransferpaper}, this offset is a result of replacing the Dirac $\delta$ distribution with a finite-width Gaussian when calculating the LDOS in a practical LDOS-based model. This offset is further unproblematic, since physical properties are related to energy differences, not absolute energy values and as such, it is important that energy predictions reproduce correct relative behaviors, which the LDOS based models achieve. 

The results shown in Fig.~\ref{fig:ionic_temperature_comparison} and \ref{fig:electronic_temperature_comparison} lead to three important observations. First, it is evident that both electronic and ionic temperature influence the calculation of the electronic total free energy, i.e., energies change considerably with increasing temperatures. Secondly, while for lower temperatures changes in ionic temperature lead to larger changes in the total free energy, electronic temperature effects become relevant at higher temperatures as well. In general, electronic effects may become dominating at high temperatures for specific observables, such as the thermal conductivity \cite{thermalconduct} (see also App.~\ref{app:al_motivation}). And finally, LDOS based models, are \textit{in principle} capable of recovering most energetic effects related to the electronic temperature, which is an important property when going to high temperature regimes.

\begin{figure}[h]
    \centering
    \includegraphics[width=0.95\columnwidth]{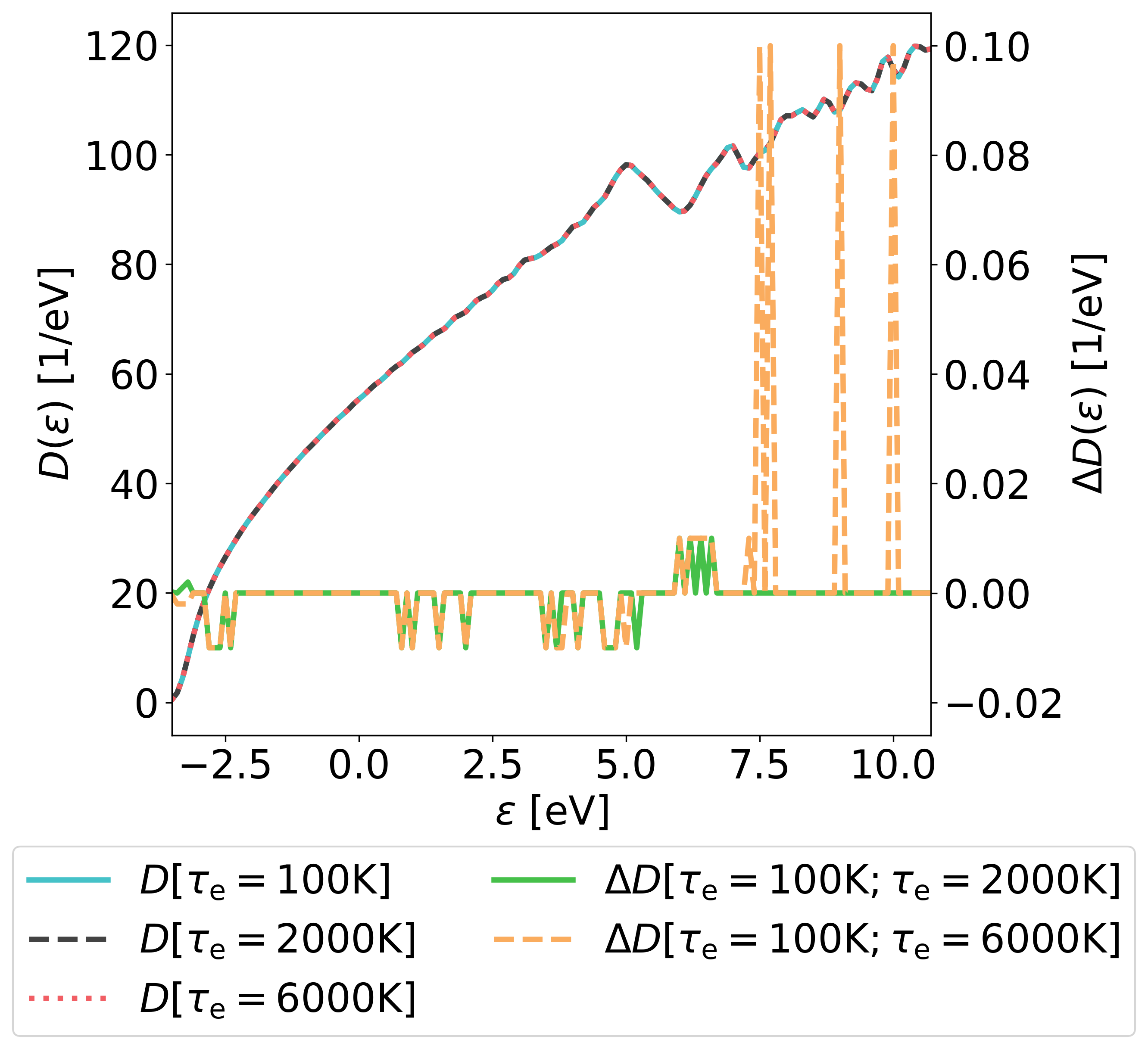}
    \caption{Density of states (DOS) for the same aluminum configuration but varying electronic temperatures, including absolute differences $\Delta D(\epsilon)$ shown in black and red.}
    \label{fig:dos_comparison}
\end{figure}

In order to comprehend why the LDOS accurately encapsulates these effects, it is crucial to recognize that the LDOS and hence the DOS exhibit only minimal variation with electronic temperature, as depicted in Fig.~\ref{fig:dos_comparison}. This can be discerned from the definitions of the LDOS and DOS in Eqs.~(\ref{eq:ldos}) and (\ref{eq:dos}). Both are dependent on the electronic temperature through the temperature dependence of the Kohn-Sham potential, which in turn depends on the electronic temperature via the corresponding Kohn-Sham orbitals and eigenvalues.
Two primary energy contributions influenced by the electronic temperature are the kinetic energy, which can be accessed through the band energy $E_b$, and the electronic entropy $S\s$, both expressed using the DOS as 
\begin{align}
    E_b = &\int d \epsilon\;  f^{\tau\electronic}(\epsilon) \epsilon D(\epsilon) \; , \\
    S\s = & -\int d\epsilon \; \big( f^{\tau\electronic}(\epsilon) \ln{\left[f^{\tau\electronic}(\epsilon)\right]}  \nonumber\\ &+ \left[1-f^{\tau\electronic}(\epsilon)\right] \ln{\left[1-f^{\tau\electronic}(\epsilon)\right]}\big) D(\epsilon) \; .
\end{align}
Thus, provided that the (L)DOS has been sampled to include enough (unoccupied) energy states at lower electronic temperatures, the LDOS can be used to accurately evaluate the energy at higher electronic temperatures. 

Therefore, the LDOS is a suitable target quantity for ML models across temperature ranges and for high-temperature regimes. The task at hand is to correctly predict the LDOS for ionic configurations sampled at varying $\tau\ionic$. Any model that performs well at this task will, by default, capture most of the relevant effects related to $\tau\electronic$. 

\section{Temperature transferable models}
\label{sec:resultsModels}
Even with LDOS-based models correctly treating effects related to electronic temperature, it remains to be determined whether ML models can correctly learn the electronic structure of ionic configurations across a temperature range. In the case of the models employed here, the ionic structure is encoded in bispectrum descriptors, and naturally, these descriptors change with the ionic temperature. Furthermore, we need to assert that models can recover the LDOS accurately enough to reproduce energy effects related to the electronic temperature, which they in theory should be able to, as shown in Fig.~\ref{fig:electronic_temperature_comparison}.

\begin{figure}[h]
    \centering
    \includegraphics[width=0.95\columnwidth]{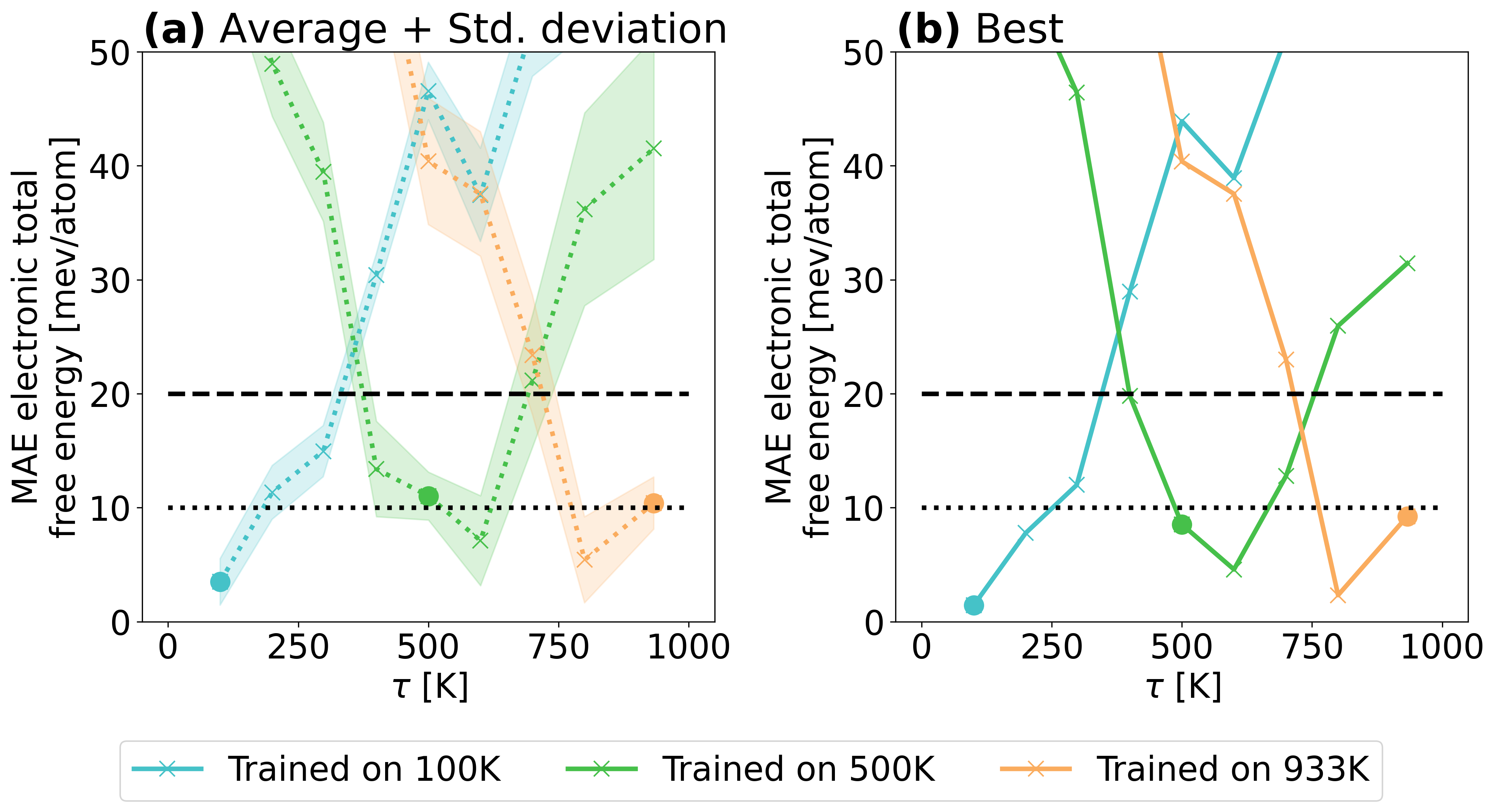}
    \caption{Performance of MALA models when trained on single temperature data. Five models with different initializations were trained for each temperature, where \textbf{(a)} shows the average and standard deviation, while \textbf{(b)} shows the performance of the best model. Training was performed using data from four ionic configurations, while validation was performed with data from one ionic configuration per temperature.}
    \label{fig:temptransfer_single_training}
\end{figure}

\begin{figure}[h]
    \centering
    \includegraphics[width=0.95\columnwidth]{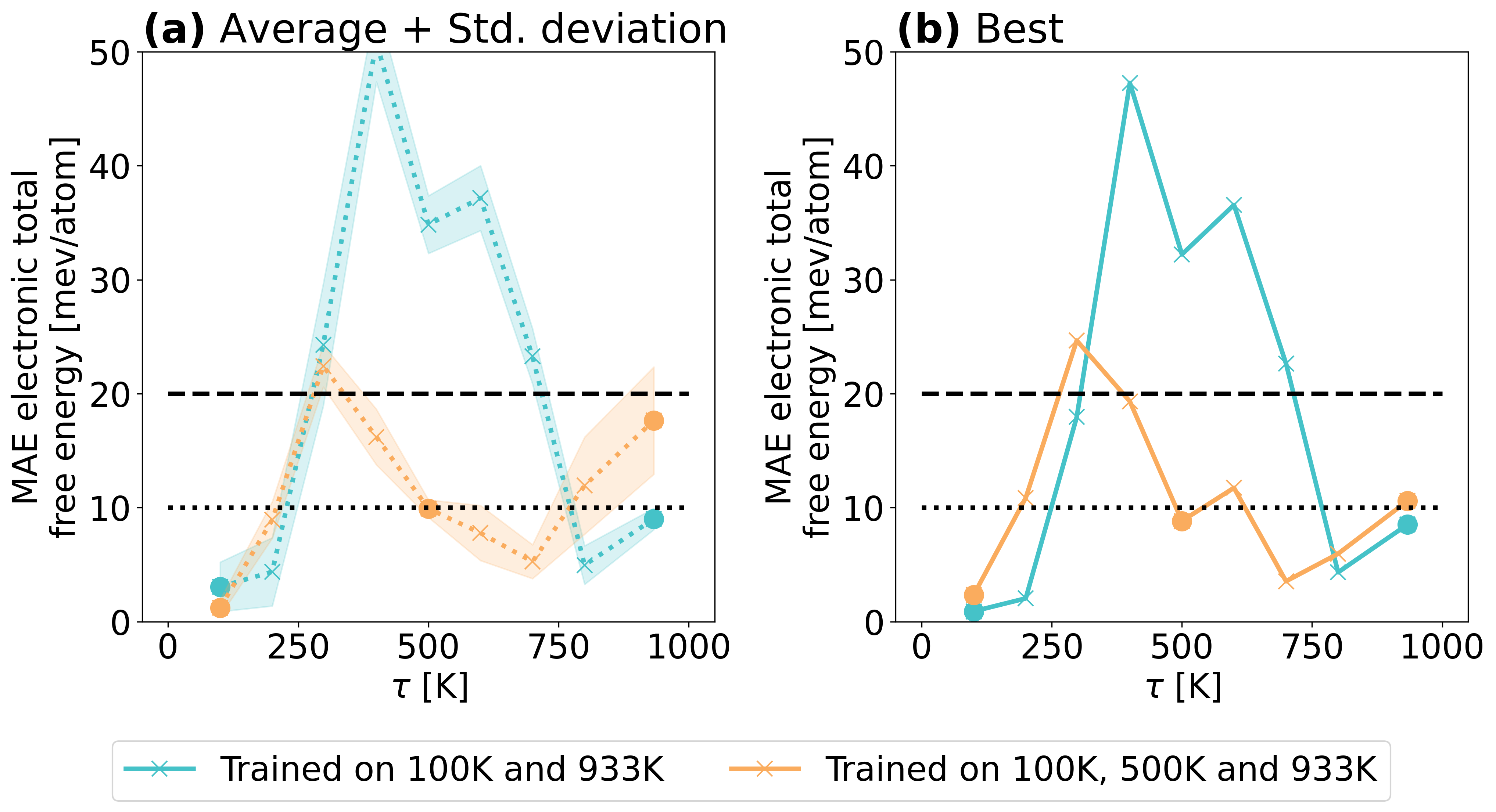}
    \caption{Performance of MALA models when trained on multiple temperatures. Two types of models were explored: one incorporating data from the beginning and end of the temperature range, and one additionally incorporating data from the middle of the temperature range, i.e., 500K. Each model was trained five times with different initializations. Panel \textbf{(a)} shows the average and standard deviation of the performances across these models, while panel \textbf{(b)} shows the performance of the best model. For each included temperature, data from four ionic configurations was used for training, while validation was performed with data from one ionic configuration.}
    \label{fig:temptransfer_hybrid_training}
\end{figure}

\begin{figure}[h]
    \centering
    \includegraphics[width=0.95\columnwidth]{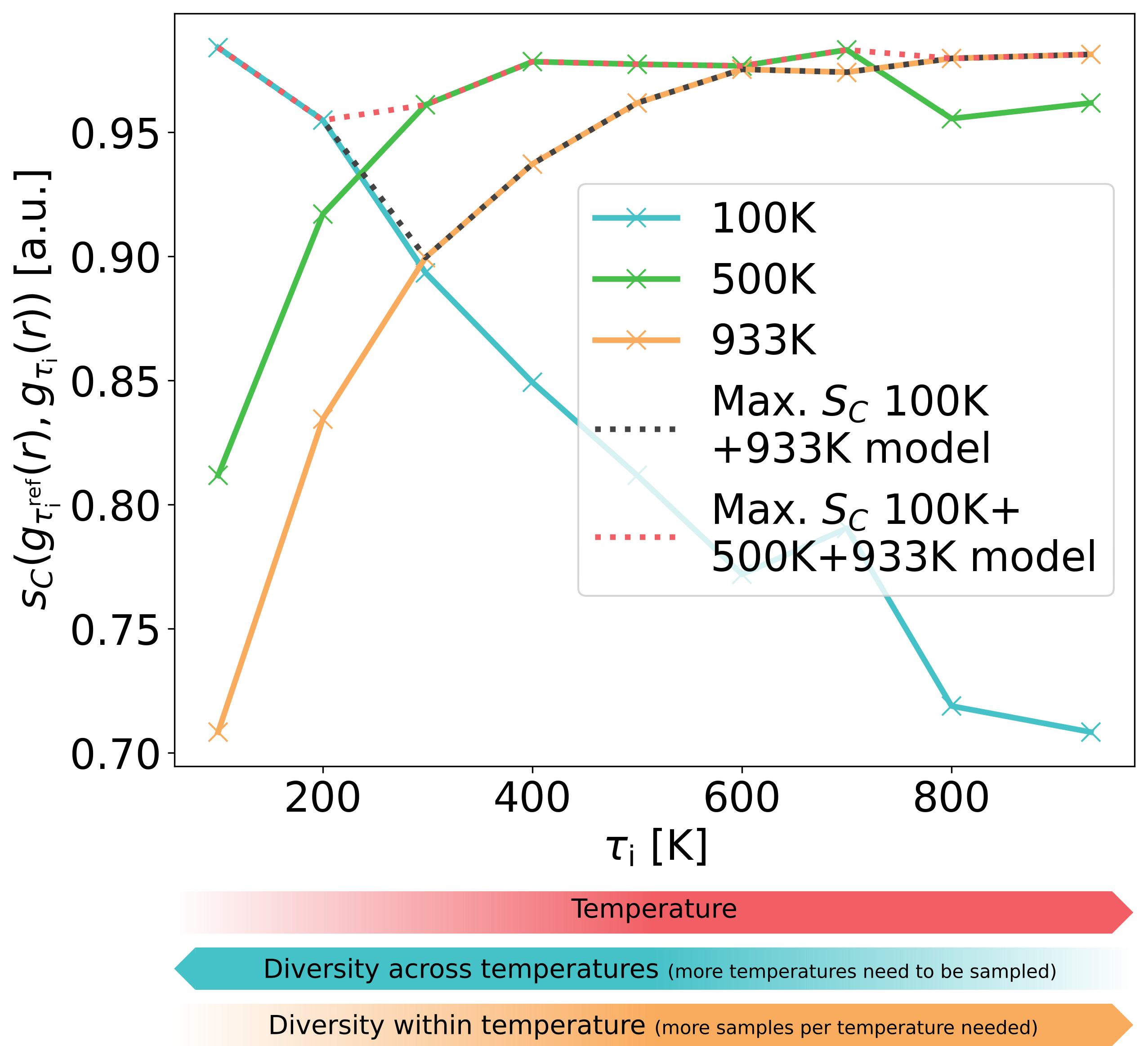}
    \caption{Cosine similarities between an arbitrary ionic configuration at reference temperatures 100K, 500K and 933K and a different configuration at all investigated ionic temperatures. The maximum cosine similarity per model is shown for the first two presented multi-temperature models, providing a metric of model performance. The pictrogram below summarizes the main result found in this analysis, i.e., configuration diversity across temperatures decreases with increasing temperature, while configuration diversity within temperature increases with increasing temperature.}
    \label{fig:rdf_temperature}
\end{figure}

\begin{figure*}[h]
    \centering
    \includegraphics[width=0.95\textwidth]{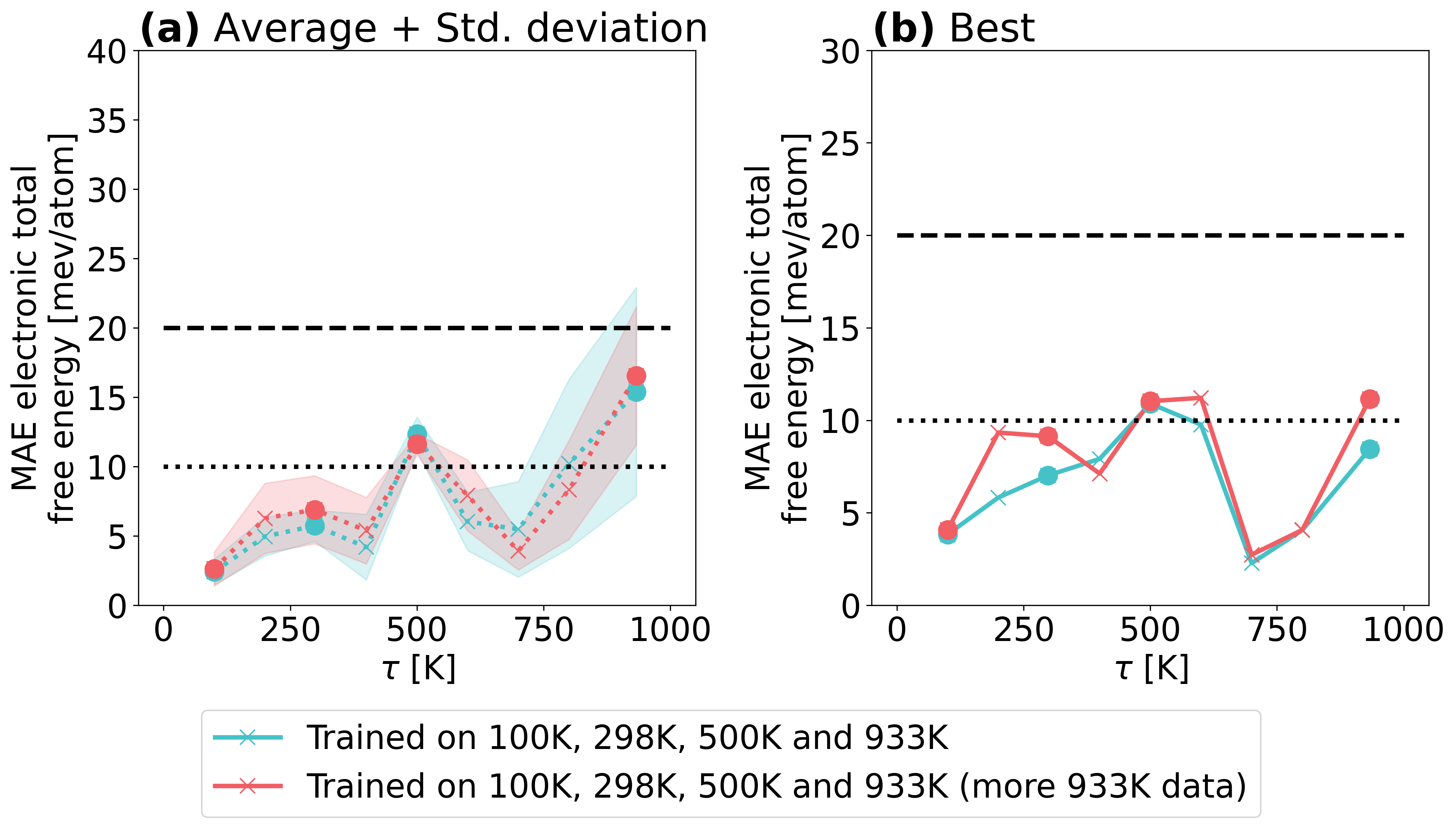}
    \caption{Performance of MALA models trained on four temperatures, i.e., 100K, 298K, 500K, and 933K. The models depicted in blue were trained with four ionic configuration and the same model architecture used for Fig.~\ref{fig:temptransfer_single_training} and Fig.~\ref{fig:temptransfer_hybrid_training}. The models depicted in red were trained with two additional ionic configurations for 933K and a slightly larger architecture (see Sec.~\ref{sec:compdetails} for more information). All models were trained five times with different initializations, and per temperature included, data from four (resp.~six in the 933K case for the models depicted in red) ionic configurations was used for training, while validation was performed with data from one ionic configuration.}
    \label{fig:temptransfer_superhybrid}
\end{figure*}

To investigate this, we construct models using different aluminum training data sets and evaluate them across a temperature range of 100K to 933K, the melting point of aluminum, in increments of 100K. In all DFT calculations used to generate the training data, the ionic and electronic temperatures were set to the same values. This is why $\tau = \tau_{\electronic} = \tau_{\ionic}$ is used to represent both the ionic and electronic temperatures in Fig.~\ref{fig:temptransfer_single_training}, Fig.~\ref{fig:temptransfer_hybrid_training}, Fig.~\ref{fig:rdf_temperature}, and Fig.~\ref{fig:temptransfer_superhybrid}.
The first step is to establish a proper baseline for this experiment. It is expected that models trained only on singular temperatures would struggle to accurately capture the electronic structure of both higher and lower temperatures. We quantified this by training models on data for either 100K, 500K, or 933K representing the beginning, middle, and end of the temperature range. 

The results are shown in Fig.~\ref{fig:temptransfer_single_training}, which displays the electronic total free energy errors for models trained on only one temperature. We trained five models per temperature to assess the robustness of the models. Both the average and standard deviation across the initializations are given, along with the model that performs best. It is evident that for all three temperature, the models become increasingly inaccurate as one moves away from the training temperature. Specifically, we seek electronic total free energy errors of below 10 meV/atom. In the best case, i.e., the models shown in Fig.~\ref{fig:temptransfer_single_training}\textbf{(b)}, this threshold is barely met for temperatures directly adjacent to the training temperature, and quickly exceeded as one moves to lower or higher temperatures. 

As mentioned above, such a behavior is expected, since NN based models usually perform poorly at extrapolation tasks, and training a model on merely one ionic temperature inevitably constitutes an extrapolation in the ionic temperature domain. In order to build transferable ML models for the electronic structure one has to take multiple temperatures into account, and the principal question is how many and which temperatures need to be incorporated into such a model in order to produce accurate results across the selected temperature range. 

There are two conceivable ways to construct such a model. One potential route is to take training data from the entire temperature range into account. However, since ML models based on NNs should perform well when used in an interpolative fashion, a different strategy is to investigate the number and proper selection of temperatures needed to train models that perform well across the temperature range. 

To this end, we have trained models using training sets with different combinations of temperatures. In all cases, the amount of training data \textit{per temperature} has been kept constant, and five models per approach have been trained with different initializations, to quantify the robustness of the method. The results are shown in Fig.~\ref{fig:temptransfer_hybrid_training}.
The computational cost for training models increases as the number of data points and the temperature range expands. We provide a brief discussion on this trade-off between accuracy and cost in Appendix~\ref{app:training_time}.

The first type of model, shown in blue in Fig.~\ref{fig:temptransfer_hybrid_training}, uses training data from the beginning and end of the temperature range. Although technically the inference across the temperature range is an interpolation task, the reported accuracies are unsatisfactory in the middle of the temperature range. As previously observed in Fig.~\ref{fig:temptransfer_single_training}, accuracies are only sufficient for 200K around the training temperature, i.e., close to temperatures observed in training, and thus the resulting models cannot make accurate predictions from around 300K to 600K. 

A obvious solution to address this problem is to include of training from the middle of the temperature range. A first attempt at this is shown in Fig.~\ref{fig:temptransfer_hybrid_training} in orange, where training data at 500K was added to the model. As expected, this substantially reduces the inference errors across the middle of the temperature range. In fact, the resulting model achieves competitive accuracy almost throughout the entire temperature range. 

Further analysis of errors reveals two main sources of error. Firstly, there seems to be a noticeable decrease in accuracy between 100K and 500K compared to temperatures above 500K. The reason for this lies in the fact that while diversity within the data set increases as one moves to higher temperatures, differences between temperatures become more subtle, and conversely, differences in ionic structure are more pronounced at lower temperatures. This can be further verified by looking at the radial distribution function (RDF) \cite{rdf1, rdf2, rdf3}, which is calculated by averaging the ion density contained in a shell of radius $[r, r+dr]$ for a cell of volume $V(r)$ and isotropic system density $\rho=N/V$, i.e.,  
\begin{equation}
    g(r) = \frac{1}{\rho\,N\,V(r)}\,\sum_{i=1}^N \sum_{\substack{j=1\\j\neq i}}^N
                \delta(r - \left\vert{\br}_i-{\br}_j\right\vert) \; .
\end{equation}
Differences in the RDF can best be analyzed through the cosine similarity, which is a standard metric in data science analysis \cite{cosine_distance_reference} and is defined as
\begin{equation}
    s_C\left( \bm{x}, \bm{y}\right) = \frac{\bm{x}\cdot \bm{y}}{\|{\bm{x}}\| \|\bm{y}\|} \label{eq:cosine_similarity} \; 
\end{equation}
for vectors $\bm{x}$ and $\bm{y}$. By interpreting the RDFs of individual temperatures as vectors of dimensionality $N_r$, where $N_r$ is the number of radii sampled for the calculation of the RDFs, one can calculate their similarities. A similarity $s_C=1$ indicates that the two RDFs are identical, while $s_C=0$ implies fully orthogonal RDFs. We have carried out such an analysis for the temperatures included in the model shown in Fig.~\ref{fig:temptransfer_hybrid_training}. The results are presented in Fig.~\ref{fig:rdf_temperature}. 

It is evident that a drop in model accuracy corresponds to areas for which none of the included configurations at the training temperatures exhibit high cosine similarities to the inference configurations. For instance, a clear lack of cosine similarity can be observed in the region between 200K and 600K when only considering 100K and 933K reference data; this corresponds to the lack of accuracy observed for the first model shown in Fig.~\ref{fig:temptransfer_hybrid_training}. Similarly, even with the inclusion of 500K training data, such a gap persists for temperatures of around 200K and 298K, which corresponds to the sudden drop in accuracy observed for the second model in Fig.~\ref{fig:temptransfer_hybrid_training}. Visualizing the combined maximum cosine similarity provides a good metric of expected model performance.

Therefore, to construct a final temperature-transferable model, we incorporate 298K data into the model training process. The resulting model is shown in Fig.~\ref{fig:temptransfer_superhybrid}, depicted in blue. We incorporate the same amount of data for each training temperature, and train the model with five different initializations to quantify the robustness of the method.

The resulting models exhibit excellent performance throughout the temperature range. For the best of the five initializations, errors consistently remain below 10 meV/atom throughout the entire test set, and on average, this threshold is only slightly exceeded. While there is a small linear increase in error when moving to higher temperatures, the models are generally capable of delivering consistent accuracy across the entire temperature range.

However, as mentioned above, a second issue remains with the models both shown in Fig.~\ref{fig:temptransfer_hybrid_training} and Fig.~\ref{fig:temptransfer_superhybrid} (blue). When considering the entire ensemble of trained models, it is apparent that the models become less robust as temperature increases. This can be attributed to the fact that while pronounced differences between temperatures play an important role for lower temperatures, a larger diversity of data points per temperature is observed at higher temperatures, necessitating consideration of more diverse ionic and electronic environments.

One way to mitigate these issues is to include more data at the problematic temperatures, such as 933K. However, this approach may exceed the information capturing capacity of the employed model for the given data. To address this, we trained models with two additional ionic configurations at 933K with the results also given in Fig.~\ref{fig:temptransfer_superhybrid}, depicted there in red. Slightly larger NNs were used to successfully train these models (see Sec.~\ref{sec:compdetails} for detailed information). The inclusion of additional data at 933K leads to an increase in the systematic model robustness across the temperature range, i.e., while the average accuracy did not change, the standard deviation decreased. There is a slight decrease in accuracy for lower temperatures, yet errors are in the same range and the 10 meV/atom threshold is only slightly exceeded. Overall, models trained with additional data at 933K demonstrate excellent accuracy and robustness across the entire temperature range.

\begin{figure*}[h]
    \centering
    \includegraphics[width=0.95\textwidth]{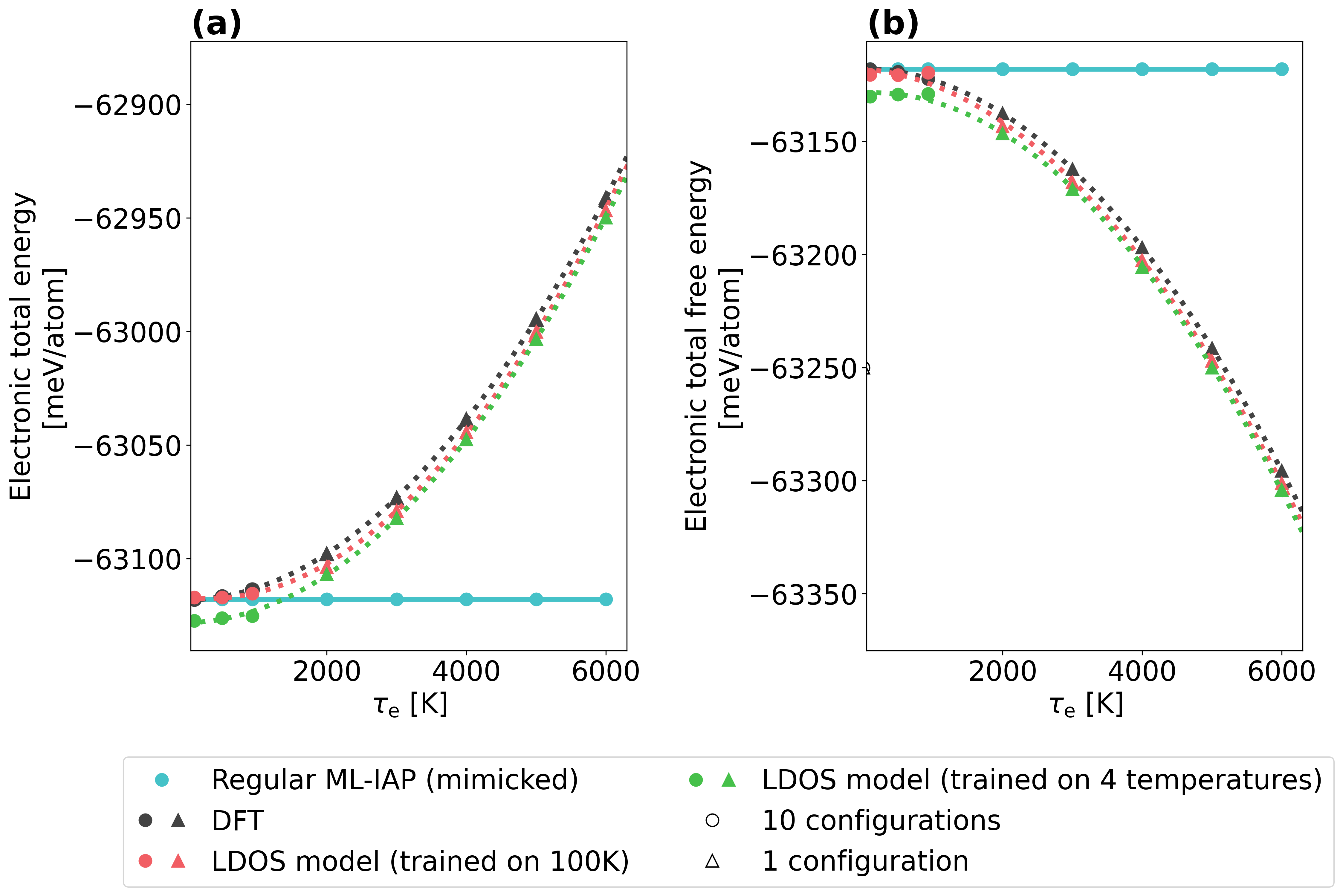}
    \caption{Comparison of MALA model inference with varying electronic temperature as compared to DFT, idealized LDOS models and ML-IAPs. The same methodology and configurations as in Fig.~\ref{fig:electronic_temperature_comparison} have been used, i.e., the ionic temperature was kept constant at 100K, ten configurations were sampled for the points denoted with circles and one for the points denoted with a triangle, while dotted lines denote a quadratic fit. The model shown in red corresponds to the 100K model shown in Fig.~\ref{fig:temptransfer_single_training}, while the model shown in green corresponds to the model trained on data from four temperatures (i.e., the model shown in red in Fig.~\ref{fig:temptransfer_superhybrid}).}
    \label{fig:ml_extrapolation}
\end{figure*}

The models shown in Fig.~\ref{fig:temptransfer_superhybrid} show good transferability across a range of ionic temperatures. As introduced in Sec.~\ref{sec:resultsTheory}, LDOS based models should further be able to accurately predict the electronic total free energy at electronic temperatures different, esp.~higher than they were originally trained on, since the (L)DOS is roughly constant with changing electronic temperature. To this end, one can calculate the total (free) energies for the same configurations as used in Fig.~\ref{fig:electronic_temperature_comparison}. In Fig.~\ref{fig:ml_extrapolation} this has been done for two MALA models discussed here, namely the model trained solely on 100K data shown in Fig.~\ref{fig:temptransfer_single_training} and the final multi-temperature model shown in Fig.~\ref{fig:temptransfer_superhybrid}. In each case, the model initialization with the best performance as shown in the respective figures have been used.

In Fig.~\ref{fig:ml_extrapolation} it can clearly be seen that overall, both MALA models reproduce the quadratic dependence of the energy with respect to the electronic temperature quite well. It should be noted that this result represents an interpolation from a pure ML perspective, since the LDOS was predicted from ionic configurations similar to those observed in training, while from a physics perspective, these results constitute an extrapolation. Due to the aforementioned weak dependence of the (L)DOS on electronic temperature, this becomes feasible. Results for two different models are shown to illustrate the subtle point that the accuracy of this temperature extrapolation depends on a model's capability of reproducing the (L)DOS for an ionic temperature of 100K. Since this ionic temperature was fixed in the results shown in Fig.~\ref{fig:ml_extrapolation}, it is evident from the results of an idealized LDOS model shown in Fig.~\ref{fig:electronic_temperature_comparison} that this must be the case. The model trained solely on 100K data thus performs slightly better at this task, almost exactly recovering the DFT total energy in the low temperature case and only slightly deviating from it in the electronic total free energy case. Conversely, the multi-temperature model, which shows generally excellent performance across the ionic temperature range, performs slightly worse in these conditions, especially in the total free energy prediction shown in Fig.~\ref{fig:ml_extrapolation}. The difference in accuracy between electronic total free and electronic total energy may be explained by the fact that for the former, the electronic entropy is factored in; for this, a DOS integration has to be performed with a numerically determined Fermi energy as explained in Ref.~\cite{malapaper}, which can lead to slight inaccuracies. 


Generally, both models recover the relative energy behavior with increasing electronic temperature quite well. Especially with respect to the multi-temperature model this is an important result, since it shows that a singular ML model can be used to predict the electronic structure of a material at varying ionic and electronic temperatures. Both temperatures may be varied independently of one another, within reasonable boundaries, yet the same model can be employed. Therefore, MALA models can replace DFT fully in this regard, since both parameters are independent inputs for DFT simulations as well. 

\section{Discussion and Outlook}
\label{sec:discussion}
In summary, our work has demonstrated the potential of ML models in replacing DFT calculations for electronic structure predictions across a range of temperatures. By targeting the LDOS, we have shown that ML models can accurately reproduce the electronic structure of ionic configurations at temperatures unseen during the training process, and are further capable of extrapolation in the electronic temperature domain. This is due to the fact that the LDOS by default recovers effects at higher electronic temperatures than it was originally calculated at. Thus, any model that is capable of predicting the LDOS at arbitrary ionic configurations will be less susceptible to errors due to changes in electronic temperature. This only holds true if the model actually reproduces an accurate LDOS for an arbitrary ionic configuration and if the LDOS has been sampled for large enough energies in the energy domain. The resulting models are capable of reproducing DFT fully within the temperature ranges for which they apply, and one may use them to model any combination of ionic and electronic temperatures. 

We have also investigated the number and selection of temperatures needed to train transferable ML models. We found that a careful selection of temperatures, aided by analyzing the RDFs of the system, can lead to highly accurate and robust models. 

While our work focusses on the simple system of aluminum in the solid phase up to the melting point, our findings provide a foundation for future investigations into more complicated systems and conditions, such as warm dense matter. In these cases, more elaborate descriptors and powerful model representation may be necessary, and we look forward to further exploration in these areas. 
Overall, our work highlights the potential of ML models in accurately predicting the electronic structure across a range of temperatures, with broad implications for materials science and beyond.

\section*{Acknowledgments}
The authors acknowledge helpful feedback from Chen Liu, SambaNova Systems, with regards to data set preparation. The authors thank Kieron Burke for motivating the scientific question addressed in this paper.
Sandia National Laboratories is a multi-mission laboratory managed and operated by National Technology and Engineering Solutions of Sandia, LLC, a wholly owned subsidiary of Honeywell International, Inc., for DOE's National Nuclear Security Administration under contract DE-NA0003525.
This paper describes objective technical results and analysis.
Any subjective views or opinions that might be expressed in the paper do not necessarily represent the views of the U.S. Department of Energy or the United States Government.
This work was
supported by the Center for Advanced Systems Understanding (CASUS) which is financed by Germany’s Federal Ministry of Education and Research (BMBF) and by the Saxon state government out of the State budget approved by the Saxon State Parliament.
The authors acknowledge the Center for Information Services and High Performance Computing [Zentrum für Informationsdienste und Hochleistungsrechnen (ZIH)] at TU Dresden for providing its facilities for high-performance calculations.

\appendix

\section{Estimation of relevant electronic temperatures for aluminium}
\label{app:al_motivation}

As discussed in Sec.~\ref{sec:resultsTheory} and \ref{sec:resultsModels}, we have investigated the behavior of the electronic total (free) energy when keeping ionic temperature fixed at relatively small values, while increasing electronic temperature. The overarching motivation for this kind of experiment are conditions in which electrons are heated to temperatures far exceeding ionic temperatures, e.g., by means of laser heating. As our models extend only to solid ionic configurations, we have selected the upper limit for electronic temperature as the temperature at which the electrons would be just hot enough to melt the material. Naturally, this is only a rough estimate based on classical formulas intended to guide the temperature range investigated. 

One can make such an estimation by first calculating the electronic thermal energy as
\begin{equation}
    E\electronic (\tau\electronic)= \frac{1}{2} \gamma \tau\electronic^2 \; . \label{eq:electronicthermalenergy}
\end{equation}
This derivation is based on the free electron model \cite{sommerfeld1928elektronentheorie}, in which the electronic volumetric heat capacity is given as $c_v = \gamma \tau\electronic$, with the material constant $\gamma$. Conversely, the ionic thermal energy above the Debye temperature (which is 433K in the low temperature limit for aluminium \cite{tari2003specific}) can be expressed as 
\begin{equation}
    E\ionic(\tau\ionic)=\frac{3}{2} N_A \kB \tau\ionic \; , \label{eq:ionicthermalenergy}
\end{equation}
according to the Dulong-Petit law \cite{dulong1819recherches}, to which, e.g., the Debye model \cite{debyemodel} reduces at high temperatures. Eq.~(\ref{eq:ionicthermalenergy}) implies a constant ionic volumetric heat capacity, which is a central assertion of the Dulong-Petit law. Based on these simple approximations one can estimate the heat energy necessary to heat electronic and ionic systems. It is further evident that \textit{via} these assumptions, the electronic contributions to the thermal conductivity dominate at high temperatures. 
To quantify the heat energy needed to melt a material, one further needs to factor in the heat of fusion $E_\mathrm{fusion}$, i.e., the energy associated with the phase change from solid to liquid phase.

With these considerations, the energy necessary to melt aluminium purely through electronic heating can be defined as 
\begin{align}
    E^\mathrm{melt}\electronic = &E\electronic (\tau^\mathrm{melt}\electronic) \nonumber \; , \\= &E\ionic(933~\mathrm{K}) + E\electronic(933~\mathrm{K}) + E_\mathrm{fusion} \; . \label{eq:meltingelectronicenergy}
\end{align}
Thus, Eq.~(\ref{eq:meltingelectronicenergy}) states that the electronic thermal energy necessary to melt initially  solid aluminium at $\tau\ionic=\tau\electronic=0~\mathrm{K}$ is equal to the sum of both ionic and electronic thermal energies at the melting point as well as the heat of fusion itself. Conceptually speaking, the external heating of the electrons would have to provide sufficient heat energy to heat both ionic and electronic systems to 933K and provide the heat of fusion. 

With the experimentally determined values $\gamma = 1.35 \cdot 10^{-3}~\mathrm{J}/(\mathrm{mol\,\mathrm{K}})$ \cite{kittel2021introduction} and $E_\mathrm{fusion} = 10431.1~\mathrm{J}/\mathrm{mol}$ \cite{heatoffusion}, one can determine $E^\mathrm{melt}\electronic$ as $E^\mathrm{melt}\electronic = 22654.8~\mathrm{J}/\mathrm{mol}$. This corresponds to a  temperature $\tau^\mathrm{melt}\electronic = 5793.3~\mathrm{K}$. 

Therefore, we show both DFT and ML results for electronic temperatures up to 6000~K and, in doing so, assert that our models are capable of reproducing the correct physics up to physically relevant temperatures.

\section{Model training times and data volume}
\label{app:training_time}

The models presented in Section~\ref{sec:resultsModels} can be evaluated in terms of their accuracy-cost trade-off, which is typical for ML models. More accurate models, such as those depicted in Fig.~\ref{fig:temptransfer_superhybrid}, generally require a larger computational effort for training. Hence, we provide the associated training times and data volume in Table~\ref{tab:timingtable}. The reported training times are averaged over all five model initializations per model type, while the data volume represents the cumulative size of the training and validation configuration data used during training.

\begin{table}
	\centering  
	\caption{Training times and data volumes for models trained in Section~\ref{sec:resultsModels}. All models were trained on a single GPU, so the reported timings in hours (h) also correspond to the computational cost in GPU-hours. Timings are averaged over all five model initializations per experiment. Data volume includes the total size of both the training and validation data sets used in the model training.}\label{tab:timingtable}		
	\begin{tabularx}{\columnwidth}{>{\hsize=.6\hsize}X >{\hsize=.2\hsize}X >{\hsize=.2\hsize}X}
	\toprule
	\textbf{Model} &  \textbf{Training} & \textbf{Data}	\\
                      &  \textbf{time [h]} & \textbf{volume [GB]}	\\
        \midrule
        Model 1                    &  3.24 & 107.50 \\
        100~K                      &       &        \\
        (Fig.~\ref{fig:temptransfer_single_training}) &  & \\
        &  &  \\ 
        Model 2                    &  2.19 & 107.50 \\
        500~K                      &       &        \\
        (Fig.~\ref{fig:temptransfer_single_training}) &  & \\
        &  &  \\ 
        Model 3                    &  2.05 & 107.50 \\
        933~K                      &	   &        \\
        (Fig.~\ref{fig:temptransfer_single_training}) &  & \\
        &  &  \\ 
        Model 4                    &  7.51 & 215.00 \\
        100~K, 933~K               &       &        \\
        (Fig.~\ref{fig:temptransfer_hybrid_training}) &  & \\
         &  &  \\ 
        Model 5                    & 15.19 & 322.51 \\
        100~K, 500 K, 933 K        &       &  \\
        (Fig.~\ref{fig:temptransfer_hybrid_training}) &	 & \\
        &  &  \\ 
        Model 6                    & 14.19 & 430.01 \\
        100~K, 298~K, 500~K, 933~K &       &        \\
        (Fig.~\ref{fig:temptransfer_superhybrid})	  &  & \\
        &  &  \\ 
        Model 7                    & 24.21 & 473.01 \\
        100~K, 298~K, 500~K, 933~K &       &        \\
        (more 933~K data, Fig.~\ref{fig:temptransfer_superhybrid}) &  & \\
        \bottomrule
    \end{tabularx}
\end{table}

The relationship between the computational cost for training a model and the number of training data points associated with a given temperature is nearly linear, with a few notable exceptions. For instance, the initial four-temperature model (Model 6) shown in Figure \ref{fig:temptransfer_superhybrid} can be trained in less time compared to the three-temperature model (Model 5) illustrated in Figure \ref{fig:temptransfer_hybrid_training}. This discrepancy arises from the use of early stopping during the training process, where training is terminated once the accuracy on the validation data set no longer improves. This indicates that a model has effectively captured all the necessary information from the training data set. A more balanced data set in Model 6 allows it to reach this point earlier than Model 5, which includes fewer temperature data points. Moreover, the four-temperature model incorporating the additional 933 K data (Model 7) requires relatively longer training times due to the utilization of a larger neural network compared to the other models listed in Table \ref{tab:timingtable}. Nevertheless, the increase in computational effort needed to tune temperature transferable models is moderate when considering the enhanced accuracy demonstrated in Fig.~\ref{fig:temptransfer_single_training}, Fig.~\ref{fig:temptransfer_hybrid_training}, and Fig.~\ref{fig:temptransfer_superhybrid}.

\end{document}